\documentclass[preprint,12pt]{elsarticle}

\usepackage{amssymb}
\usepackage{amsmath}
\usepackage{booktabs}
\usepackage{lineno}
\usepackage{algorithm}
\usepackage{algpseudocode}
\usepackage{bm}
\usepackage[hidelinks]{hyperref}

\graphicspath{{./}}
\journal{Acta Astronautica}

\newcommand{\NIS}{\mathrm{NIS}}
\newcommand{\Thac}{T_{\text{HAC-DANIS}}}
\newcommand{\E}{\mathbb{E}}
\DeclareMathOperator{\Var}{Var}
\DeclareMathOperator{\Cov}{Cov}

\begin{document}

\begin{frontmatter}

\title{A heteroskedasticity and autocorrelation consistent test for process noise calibration of Kalman filters in multi-sensor orbit determination\tnoteref{abbrev}}
\tnotetext[abbrev]{Abbreviations: ANEES, average normalised estimation error squared; ANIS, average normalised innovation squared; DANIS, dimensionless averaged normalised innovation squared; HAC, heteroskedasticity and autocorrelation consistent; NEES, normalised estimation error squared; NIS, normalised innovation squared.}

\author[us]{Antonio Malaver\corref{cor1}}
\ead{amalaver@us.es}
\author[us]{Jose M. Montilla}
\ead{jmontillag@us.es}
\author[us]{Rafael Vazquez}
\ead{rvazquez1@us.es}
\cortext[cor1]{Corresponding author.}

\affiliation[us]{organization={Department of Aerospace Engineering, Universidad de Sevilla},
  addressline={Camino de los Descubrimientos s/n},
  city={Sevilla},
  postcode={41092},
  country={Spain}}

\begin{abstract}
Process noise calibration is a central problem in sequential orbit determination. If the process noise covariance is too small the filter is overconfident, and if it is too large it is overconservative. In both cases the estimation error grows and the reported covariance stops describing it, which affects downstream products such as conjunction assessment or re-entry prediction. No widely adopted method diagnoses the calibration in multi-sensor settings, where measurements of different type and dimension must enter a single consistency metric. This work presents the Heteroskedasticity and Autocorrelation Consistent Dimensionless Averaged Normalised Innovation Squared (HAC-DANIS) test, which evaluates process noise calibration from the innovation sequence of the filter, without ground truth. The normalised innovation squared (NIS) of each measurement is divided by its dimension, which puts sensors of any dimension on a common scale, and the standard error of the averaged sequence is obtained with the heteroskedasticity and autocorrelation consistent (HAC) Newey--West estimator, which accounts for the sensor-dependent variance of the sequence and for the serial correlation left by a miscalibrated filter or an imperfect dynamical model. Under mixing and moment conditions the statistic is asymptotically standard normal for a calibrated filter, and for a miscalibrated one its sign indicates the direction of the miscalibration. The autocorrelation function of the normalised sequence complements the test with a characterisation of the mistuning regime. Applied to an unscented Kalman filter processing simulated laser, telescope and radar observations of four satellites in low Earth orbit, the test rejects the eight miscalibrated process noise configurations with the sign of their miscalibration and does not reject the four calibrated with it.
\end{abstract}


\end{frontmatter}


\section{Introduction}\label{sec:intro}

Sequential orbit determination with an unscented Kalman filter (UKF) \cite{julier1997new} requires a process noise covariance $\boldsymbol{Q}$ that describes the dynamical model errors the filter cannot represent. If $\boldsymbol{Q}$ is too small, the filter is overconfident: it reports a covariance smaller than its actual errors, gives too little weight to the measurements and can diverge. If $\boldsymbol{Q}$ is too large, the filter is overconservative: it reports inflated uncertainty and gives too little weight to its own prediction. In both cases the reported covariance does not describe the estimation error, and every product computed from it, such as a collision probability or a re-entry prediction, inherits the discrepancy. Space surveillance and tracking networks pose a further difficulty: a single filter processes measurements of different type and dimension. For example, the Spanish Space Surveillance and Tracking (S3T) network \cite{gomez2017description} combines satellite laser ranging, optical telescopes and surveillance radars.

The agreement between the reported covariance and the actual error is classically tested in state space with the normalised estimation error squared (NEES), which follows a chi-square distribution with as many degrees of freedom as the state dimension when the filter is consistent, and with its average over independent Monte Carlo runs, the average NEES (ANEES) \cite{bar2001estimation}. Both need the true state, which only a simulation provides. The filter--smoother test of McReynolds \cite{mcreynolds1990fixed} uses the smoothed estimate as the reference instead and checks whether the difference between the filtered and the smoothed estimates is consistent with the difference between their covariances. It can be evaluated, however, only after the smoother has processed the complete measurement sequence.

A separate body of work estimates or adapts $\boldsymbol{Q}$ instead of testing it. Covariance matching \cite{myers1976adaptive} estimates $\boldsymbol{Q}$ online from the sample statistics of the state corrections over a window of past epochs, assumes that the noise statistics are constant over that window, and needs ad hoc corrections to keep the estimate positive semi-definite. Bayesian \cite{sarkka2009recursive}, maximum likelihood \cite{shumway1982approach} and correlation \cite{odelson2006new} methods treat the noise covariances as unknown parameters of the state-space model and estimate them from the data, with different computational costs, convergence guarantees and ranges of validity for nonlinear time-varying systems. Dun\'{\i}k et al.\ \cite{dunik2017noise} survey and compare these four families of methods. State noise compensation and dynamic model compensation \cite{tapley2004statistical} represent the unmodelled accelerations as white noise or as a first-order Gauss--Markov process whose parameters are tuned offline, so the tuning has to be repeated when the dynamical environment changes. In orbit determination, Stacey and D'Amico \cite{stacey2021adaptive} estimated the state noise compensation and dynamic model compensation parameters online, by fitting the intensity of the unmodelled accelerations so that the resulting $\boldsymbol{Q}$ matches the covariance matching estimate. These methods produce a value of $\boldsymbol{Q}$; they do not test whether a given $\boldsymbol{Q}$, however obtained, is consistent with the innovations of sensors of different dimension under serial dependence, which is the task of this paper.

Tests built on the innovations need no reference trajectory. When the filter is consistent, the normalised innovation squared (NIS) of each measurement follows a chi-square distribution with as many degrees of freedom as the measurement dimension, and its time average, the average NIS (ANIS), and its sum over a sequence of measurements give tests with more power than a single epoch \cite{bar2001estimation}. An optimal filter also produces white innovations, and the whiteness test of Mehra \cite{mehra1970identification} checks their sample autocorrelation.

The normalised squares of the innovations have a variance that depends on the dimension of the sensor that produced them, so their sequence is heteroskedastic, that is, its variance is not constant, even for a consistent filter. Furthermore, when the process noise is miscalibrated, the dynamical model is imperfect, or both happen at the same time, the innovations of a filter that tracks a satellite become serially correlated as well. Inference on the mean of such a sequence is a standard problem in econometrics, where heteroskedasticity and autocorrelation consistent (HAC) estimators give the standard error of a sample mean without a model of the dependence \cite{newey1987simple}. To our knowledge, HAC estimators have not been applied to test the consistency of Kalman filters.

This paper presents the Heteroskedasticity and Autocorrelation Consistent Dimensionless Averaged Normalised Innovation Squared (HAC-DANIS) test, which checks the process noise calibration of a sequential orbit determination filter that processes measurements of different type and dimension. The test rests on two ideas. First, the NIS of every measurement is divided by its dimension and centred, so that laser, telescope and radar measurements enter a single sequence that has zero mean for a consistent filter and in which every measurement carries the same weight. Second, the standard error of the average of that sequence is computed with the Newey--West HAC estimator, so the test remains valid when the sequence is heteroskedastic, as it always is with sensors of different dimension, and serially correlated, as it is when the process noise is miscalibrated or the dynamical model imperfect. The test statistic is asymptotically standard normal (zero mean, unit variance) when the filter is consistent, and its sign shows whether the process noise is too small or too large. The test needs neither the true trajectory nor a smoother. It uses only the innovations that the filter already produces, so it applies to real tracking data and can be followed as the measurements arrive. The autocorrelation function of the normalised sequence complements the test with a description of the dependence.

The test is validated in simulation with a UKF that processes laser, telescope and radar measurements, of dimension one, two and four, of four satellites in low Earth orbit (LEO) in two dynamical regimes: the geodetic sphere Starlette and the Envisat spacecraft at about $800$~km, where gravitational perturbations dominate, and the ISS and Tiangong stations at about $400$~km, where atmospheric drag does. As in operations, the dynamical model of the filter is imperfect: it uses a reduced force model, while the reference trajectory is propagated with a high-fidelity one. The measurement noise model is exact, so any inconsistency comes from the process noise. Each satellite is processed with a deliberately overconfident and a deliberately overconservative process noise, and with a calibrated one selected with the test itself, twelve scenarios in all. The test rejects the eight miscalibrated configurations with the correct sign, and the calibrated process noise gives the lowest root mean square position error of the three configurations for three of the four satellites. In the Starlette scenarios, the autocorrelation function of the normalised sequence shows a distinct pattern in each regime.

This paper is organised as follows: Section~\ref{sec:ukf} summarises the UKF and the process noise model, Section~\ref{sec:hacdanis} develops the HAC-DANIS test, Section~\ref{sec:setup} describes the simulation setup, Section~\ref{sec:results} presents the results and Section~\ref{sec:conclusions} gives the conclusions.
 \section{Unscented Kalman filter for orbit determination}\label{sec:ukf}

In this section the estimator is set up. The state, the reference frames and the dynamical and measurement models are introduced first, then the prediction and correction steps of the unscented Kalman filter and the properties of its innovations, and finally the process noise model whose calibration the test examines.

\subsection{System model}\label{sec:system}

The filter considered is discrete: it advances from one measurement epoch to the next, and a step consists of the propagation of the state and its covariance from the epoch $t_{k-1}$ of the previous measurement to the epoch $t_k$ of the current one, followed by the update with that measurement. The spacecraft state collects the Cartesian position and velocity in the Earth-centred inertial frame EME2000 and the two force coefficients estimated with them, the drag coefficient $C_D$ and the reflectivity coefficient $C_R$; therefore, the state dimension is $n = 8$ (Section~\ref{sec:estimation}). Besides the inertial frame, the local tangential, normal and cross-track (TNW) frame is used to describe the process noise: $T$ points along the velocity, $W$ along the orbital angular momentum and $N = W \times T$ lies in the orbital plane, towards the nadir for a near-circular orbit. The spacecraft state $\boldsymbol{x}_k \in \mathbb{R}^n$ at epoch $t_k$ evolves under nonlinear dynamics with additive process noise,
\begin{equation}
  \boldsymbol{x}_{k+1} = \mathbf{f}(\boldsymbol{x}_k) + \mathbf{w}_k, \qquad \mathbf{w}_k \sim \mathcal{N}(\mathbf{0}, \boldsymbol{Q}_k), \label{eq:dyn}
\end{equation}
where $\mathbf{f}$ is the numerical integration of the equations of motion over one step, with the force model of Section~\ref{sec:satellites}, and the process noise covariance $\boldsymbol{Q}_k$ accounts for the unmodelled accelerations, that is, the difference between the true accelerations and those the model includes. A sensor observes the state through
\begin{equation}
  \mathbf{z}_k = \mathbf{h}(\boldsymbol{x}_k) + \mathbf{v}_k, \qquad \mathbf{v}_k \sim \mathcal{N}(\mathbf{0}, \boldsymbol{R}_k), \label{eq:meas}
\end{equation}
with measurement model $\mathbf{h}$ and measurement noise covariance $\boldsymbol{R}_k$; the noises $\mathbf{w}_k$ and $\mathbf{v}_k$ are white and mutually independent. The dimension $m_k$ of $\mathbf{z}_k$ depends on the sensor. The sensors of this paper give $m_k = 1$ for a laser range, $m_k = 2$ for a pair of optical angles and $m_k = 4$ for a radar measurement of range, range rate, azimuth and elevation.

\subsection{Prediction and correction}\label{sec:ukfsteps}

The unscented transform \cite{julier1997new,wan2000unscented} represents a random vector by $2n+1$ deterministic sigma points that reproduce its mean and covariance. Propagating the points through a nonlinear function and recombining them with fixed weights gives the mean and covariance of the transformed vector to second order of its Taylor expansion, to third order for Gaussian inputs, without the Jacobians that the extended Kalman filter needs. The UKF applies the transform to the prediction and to the measurement update. Given the posterior mean $\hat{\boldsymbol{x}}_{k-1}$ and covariance $\boldsymbol{P}_{k-1}$ at $t_{k-1}$, the UKF  forms $2n+1$ sigma points,
\begin{equation}
  \boldsymbol{\chi}^{(0)}_{k-1} = \hat{\boldsymbol{x}}_{k-1}, \qquad \boldsymbol{\chi}^{(i)}_{k-1} = \hat{\boldsymbol{x}}_{k-1} \pm \Bigl(\sqrt{(n+\lambda)\,\boldsymbol{P}_{k-1}}\Bigr)_i, \quad i = 1, \dots, n, \label{eq:sigma}
\end{equation}
where $\lambda = \alpha^2(n+\kappa) - n$ is set by the tuning constants $\alpha$ and $\kappa$, the upper and lower signs correspond to the sigma points indexed by $i$ and $(i + n)$, respectively, and $(\cdot)_i$ is the $i$-th column of the Cholesky factor. Each point is propagated through the dynamics, $\boldsymbol{\chi}^{(i)}_{k|k-1} = \mathbf{f}(\boldsymbol{\chi}^{(i)}_{k-1})$, and the prior mean and covariance at $t_k$ are recovered from the weighted points,
\begin{gather}
  \hat{\boldsymbol{x}}_k^- = \sum_{i=0}^{2n} W_i^{(m)}\boldsymbol{\chi}^{(i)}_{k|k-1}, \label{eq:prior}\\
  \boldsymbol{P}_k^- = \sum_{i=0}^{2n} W_i^{(c)}\bigl(\boldsymbol{\chi}^{(i)}_{k|k-1} - \hat{\boldsymbol{x}}_k^-\bigr)\bigl(\boldsymbol{\chi}^{(i)}_{k|k-1} - \hat{\boldsymbol{x}}_k^-\bigr)^{\top} + \boldsymbol{Q}_{k-1}, \label{eq:priorcov}
\end{gather}
with the mean and covariance weights $W_i^{(m)}$ and $W_i^{(c)}$ of the unscented transform, which depend on $\alpha$, $\beta$ and $\kappa$ \cite{wan2000unscented}. The process noise enters the prior covariance only, so when the measurement $\mathbf{z}_k$ arrives at $t_k$ a second set of sigma points is drawn from the prior, $\boldsymbol{\chi}^{-(0)}_{k} = \hat{\boldsymbol{x}}_k^-$ and $\boldsymbol{\chi}^{-(i)}_{k} = \hat{\boldsymbol{x}}_k^- \pm (\sqrt{(n+\lambda)\,\boldsymbol{P}_k^-})_i$, and it is this set that the measurement model maps, $\boldsymbol{\zeta}^{(i)}_k = \mathbf{h}(\boldsymbol{\chi}^{-(i)}_{k})$. Without this second draw the predicted measurements would not see $\boldsymbol{Q}_{k-1}$, and the innovation covariance would lack the term through which the process noise reaches the innovations and the test of Section~\ref{sec:hacdanis}. The mapped points give the predicted measurement, the innovation, the innovation covariance and the state--measurement cross-covariance,
\begin{gather}
  \hat{\mathbf{z}}_k = \sum_{i=0}^{2n} W_i^{(m)}\boldsymbol{\zeta}^{(i)}_k, \qquad \boldsymbol{\nu}_k = \mathbf{z}_k - \hat{\mathbf{z}}_k, \label{eq:innov}\\
  \boldsymbol{S}_k = \sum_{i=0}^{2n} W_i^{(c)}\bigl(\boldsymbol{\zeta}^{(i)}_k - \hat{\mathbf{z}}_k\bigr)\bigl(\boldsymbol{\zeta}^{(i)}_k - \hat{\mathbf{z}}_k\bigr)^{\top} + \boldsymbol{R}_k, \label{eq:S}\\
  \boldsymbol{C}_k = \sum_{i=0}^{2n} W_i^{(c)}\bigl(\boldsymbol{\chi}^{-(i)}_{k} - \hat{\boldsymbol{x}}_k^-\bigr)\bigl(\boldsymbol{\zeta}^{(i)}_k - \hat{\mathbf{z}}_k\bigr)^{\top}. \label{eq:C}
\end{gather}
The Kalman gain $\boldsymbol{\mathcal{K}}_k = \boldsymbol{C}_k\boldsymbol{S}_k^{-1}$ updates the state and its covariance,
\begin{equation}
  \hat{\boldsymbol{x}}_k = \hat{\boldsymbol{x}}_k^- + \boldsymbol{\mathcal{K}}_k\boldsymbol{\nu}_k, \qquad \boldsymbol{P}_k = \boldsymbol{P}_k^- - \boldsymbol{\mathcal{K}}_k\boldsymbol{S}_k\boldsymbol{\mathcal{K}}_k^{\top}. \label{eq:post}
\end{equation}
This is the form implemented in the Hipparchus library, on which the unscented Kalman filter we use in Orekit is built (Section~\ref{sec:estimation}).

When the dynamical and measurement models are correct, with unbiased measurements and the correct $\boldsymbol{Q}_k$ and $\boldsymbol{R}_k$, the filter is optimal: its reported covariance is the covariance of its error (that is, it is consistent), and the innovations are zero-mean Gaussian vectors with covariance $\boldsymbol{S}_k$, independent across epochs \cite{bar2001estimation}. This holds exactly for the linear Kalman filter and, for the UKF, to the accuracy of the unscented transform. Section~\ref{sec:hacdanis} builds the test on the mean and the covariance of the innovations and does not assume their independence.

\subsection{Process noise model}\label{sec:qmodel}

The process noise covariance is defined in the TNW frame of Section~\ref{sec:system}, where each axis collects a physically distinct group of perturbations, and is taken as diagonal,
\begin{equation}
  \boldsymbol{Q}(\Delta t) = \operatorname{diag}\bigl(\sigma^{(p)\,2}_T,\, \sigma^{(p)\,2}_N,\, \sigma^{(p)\,2}_W,\, \sigma^{(v)\,2}_T,\, \sigma^{(v)\,2}_N,\, \sigma^{(v)\,2}_W\bigr), \label{eq:Qdiag}
\end{equation}
where the superscripts $(p)$ and $(v)$ denote the position and velocity components along each axis and $\sigma$ is the standard deviation of the noise accumulated over a prediction interval $\Delta t$. Following Duncan and Long \cite{duncan2006realistic}, who fit polynomials in the prediction time to the growth of the $1\sigma$ prediction errors, each standard deviation is a polynomial in $\Delta t$, for each axis $x \in \{T, N, W\}$,
\begin{equation}
  \sigma^{(p)}_{x}(\Delta t) = a_{0,x} + a_{1,x}\,\Delta t + a_{2,x}\,\Delta t^2, \quad \sigma^{(v)}_{x}(\Delta t) = b_{0,x} + b_{1,x}\,\Delta t. \label{eq:Qpoly}
\end{equation}
The interval $\Delta t_k = t_k - t_{k-1}$ is the time between consecutive updates, so a prediction of a few seconds within a pass adds little process noise and a prediction across the gap between two passes adds much more. A pass is the interval, minutes long, during which a sensor sees the satellite above its elevation mask and delivers measurements at a fixed cadence, and the passes of a sensor are separated by gaps of hours. The filter squares the polynomials evaluated at $\Delta t_k$, forms the diagonal matrix of Eq.~\eqref{eq:Qdiag} in the TNW frame and rotates it to the inertial frame, which gives the matrix $\boldsymbol{Q}_{k-1}$ of Eq.~\eqref{eq:priorcov}. The along-track coefficients are larger than the normal and cross-track ones, because the uncertainty in the direction of motion is the dominant one. The quadratic and linear terms have the form that an unmodelled acceleration gives to the position and velocity errors: an acceleration $a$ that acts over $\Delta t$ displaces the position by $a\Delta t^2/2$ and the velocity by $a\Delta t$, so $a_{2,x}$ and $b_{1,x}$ set the level of the process noise and are the coefficients that the configurations of Section~\ref{sec:qconfigs} vary. The linear coefficient of the position polynomial is $a_{1,x} = 0$, and the constant terms $a_{0,x}$ and $b_{0,x}$ are small floors that keep the process noise of a short step away from zero. The two force coefficients carry a constant process noise, small enough only to let their estimates move (Section~\ref{sec:estimation}).
 \section{HAC-DANIS consistency test}\label{sec:hacdanis}

In this section the test is built from the innovations. The normalised innovation squared and its classical aggregation are recalled first, and their limitations in a multi-sensor network motivate the dimensionless statistic and the hypothesis it examines. The standard error of the statistic is then obtained with the Newey--West estimator, the decision rule and the computation are summarised in Algorithm~\ref{alg:hacdanis}, and the autocorrelation function of the normalised sequence is introduced as a complement to the test.

\subsection{The normalised innovation squared and its aggregation}\label{sec:nis}

A filter is consistent when its estimation error has zero mean and covariance equal to the reported $\boldsymbol{P}_k$. The error is not observable outside a simulation, so the tests of this section use the counterpart of that statement in measurement space: when the filter is consistent, the innovations are zero-mean Gaussian with covariance $\boldsymbol{S}_k$, and when its model is also correct they are independent across epochs (Section~\ref{sec:ukfsteps}). The hypothesis tested below is the first part of that statement, that the innovations have on average the size the filter predicts for them. The independence is not assumed, for the reasons given in Sections~\ref{sec:danis} and \ref{sec:hac}. The normalised innovation squared (NIS) measures each innovation against its predicted covariance,
\begin{equation}
  \NIS_k = \boldsymbol{\nu}_k^{\top}\boldsymbol{S}_k^{-1}\boldsymbol{\nu}_k. \label{eq:nis}
\end{equation}
For a consistent filter, $\NIS_k$ follows a $\chi^2_{m_k}$ distribution \cite{bar2001estimation}, with $\E[\NIS_k] = m_k$ and $\Var(\NIS_k) = 2m_k$. Both moments depend on the measurement dimension, so NIS values from sensors of different dimension are not on a common scale.

The classical test \cite{bar2001estimation} aggregates the sequence into $\mathcal{S} = \sum_{k=1}^{N}\NIS_k$, which follows a $\chi^2_{\mathcal{D}}$ distribution with $\mathcal{D} = \sum_{k=1}^{N} m_k$ when the $\NIS_k$ are independent, and accepts consistency when $\mathcal{S}$ lies between two quantiles of that distribution. In a multi-sensor setting the test has two limitations. The $\chi^2_{\mathcal{D}}$ reference requires independence across epochs, which a miscalibrated filter does not provide (Section~\ref{sec:hac}). And each epoch contributes in proportion to its dimension $m_k$, so a radar epoch counts four times a laser epoch, and a miscalibration that affects mainly the low-dimensional sensors can be masked. As an example, twenty radar measurements with $\NIS = 4.0$, at their expectation, and twenty laser measurements with $\NIS = 0.2$, one fifth of their expectation, give $\mathcal{S} = 84$ with $\mathcal{D} = 100$, inside the $95\%$ acceptance region $[74.2, 129.6]$ of $\chi^2_{100}$ ($p = 0.25$): the test does not reject although every laser measurement lies far from its expected value.

\subsection{The DANIS statistic}\label{sec:danis}

Dividing each NIS value by its dimension and centring it gives
\begin{equation}
  d_k = \frac{\NIS_k}{m_k} - 1, \label{eq:dk}
\end{equation}
with $\E[d_k] = 0$ and $\Var(d_k) = 2/m_k$ for a consistent filter, whatever the sensor. The dimensionless averaged normalised innovation squared (DANIS) is the sample mean
\begin{equation}
  \bar{d} = \frac{1}{N}\sum_{k=1}^{N} d_k, \label{eq:dbar}
\end{equation}
whose expected value for a consistent filter is zero for any $N$ and any mix of sensors, and to which every epoch contributes with the same weight. The rest of the paper assumes that the measurements are unbiased and that their noise covariance $\boldsymbol{R}_k$ is correct, as in every scenario of Section~\ref{sec:setup}, so that any departure of the innovations from their predicted size comes from the process noise. A positive mean of $d_k$ then indicates an overconfident filter, with $\boldsymbol{Q}_k$ too small, and a negative mean an overconservative one, with $\boldsymbol{Q}_k$ too large. Whether the observed $\bar{d}$ differs significantly from zero is decided by the test of Section~\ref{sec:decision}.

The hypothesis the test examines is that the normalised sequence has zero mean,
\begin{equation}
  H_0:\ \E[d_k] = 0, \qquad k = 1, \dots, N. \label{eq:H0}
\end{equation}
The classical NIS test of Section~\ref{sec:nis} examines the same hypothesis, $\E[\NIS_k] = m_k$ at every epoch, but it needs the whiteness of the innovations as an assumption rather than as something it verifies: its $\chi^2_{\mathcal{D}}$ reference holds only for independent $\NIS_k$, so when the innovations are correlated the null hypothesis of the test is still correct but it answers it with the wrong error bars. HAC-DANIS asks the same question, gives every measurement the same weight whatever its dimension, and assumes nothing about whiteness. White innovations are a property of the optimal filter (Section~\ref{sec:ukfsteps}). A filter whose reported covariance is the covariance of its estimation error is consistent but need not be optimal. The three properties are nested: an optimal filter is consistent, and a consistent filter satisfies $H_0$, but neither implication reverses. $H_0$ does not require the independence of the innovations, and the standard error of Section~\ref{sec:hac} is valid whether the $d_k$ are independent or not. In either form, $H_0$ is a necessary condition for consistency but not a sufficient one; if $H_0$ is rejected we can conclude the filter is not consistent. A non-rejection is compatible with a reported covariance that describes the size of the innovations on average and does not imply that the filter is optimal: departures of opposite sign in different sensors or directions can cancel in $\bar{d}$, and the per-sensor and directional versions of the statistic that would see them are left for future work. Under the alternative $H_1$ the mean of $d_k$ is $\mu_k \neq 0$, and it need not be the same at every epoch, since the size of the innovations relative to $\boldsymbol{S}_k$ may change along the run. The quantity that $\bar{d}$ estimates is then the average $\mu_N = \frac{1}{N}\sum_{k=1}^{N}\mu_k$, and the structure of $\mu_k$ for a miscalibrated filter is left for future work.

The classical aggregate of Section~\ref{sec:nis} is the other weighting of the same quantities: $\mathcal{S} = \mathcal{D} + \sum_{k=1}^{N} m_k d_k$, so $\mathcal{S}/\mathcal{D} - 1$ is the mean of the $d_k$ weighted by the measurement dimensions. In the example of Section~\ref{sec:nis} that weighted mean is $-0.16$ and $\bar{d} = -0.40$: with equal weights the twenty laser values, each at $d_k = -0.8$, are no longer masked by the twenty radar values at $d_k = 0$. The standard error of $\bar{d}$ can be computed using Eq.~\eqref{eq:naive} of Section~\ref{sec:hac} as the $d_k$ are independent, which gives $0.18$, and the equal-weight statistic rejects the same data with $p = 0.024$.

\subsection{Standard error under heteroskedasticity and autocorrelation}\label{sec:hac}

If the $d_k$ were independent, with the variance $2/m_k$ of a consistent filter, the variance of $\bar{d}$ would be
\begin{equation}
  \Var(\bar{d}) = \frac{1}{N^2}\sum_{k=1}^{N}\frac{2}{m_k}, \label{eq:naive}
\end{equation}
and a $z$-test based on the central limit theorem would apply. The sequence is heteroskedastic, that is, its variance is not constant along the sequence, even for a consistent filter, because $\Var(d_k) = 2/m_k$ changes with the sensor, and Eq.~\eqref{eq:naive} carries these different variances. For the optimal filter, whose innovations are independent (Section~\ref{sec:ukfsteps}) so that the $d_k$ are uncorrelated, it is exact. It stops being exact in two situations, which could happen at the same time, and the test has to be able to deal with both. The first is the miscalibrated filter, whose detection is the purpose of the test. Its innovations no longer have covariance $\boldsymbol{S}_k$, so the variance of each $d_k$ differs from $2/m_k$. Consecutive measurements also share prediction errors that the filter does not remove, so the $d_k$ become serially correlated and the cross-covariances that Eq.~\eqref{eq:naive} omits no longer vanish. Both the individual variances and the cross-covariances of the $d_k$ change, and a standard error that keeps Eq.~\eqref{eq:naive} is wrong: it misstates the width of the interval for the mean of $d_k$, which Section~\ref{sec:decision} reports as the size of the inconsistency, and the power of the test near its threshold. The second situation is due to a mismatch in the dynamical models. The process noise of Eq.~\eqref{eq:dyn} is white in time, so it can match the size of the dynamical model error but not its temporal structure, and in orbit determination that error could be coloured: a mismodelled atmospheric density, for instance, pulls the prediction in the same direction over a whole pass (Section~\ref{sec:satellites}), and the drag coefficient that the filter estimates with a small process noise of its own (Section~\ref{sec:estimation}) absorbs its mean scale but not its variation. The best-tuned filter then has innovations of the predicted size on average, which is $H_0$ of Eq.~\eqref{eq:H0}, but not independent ones, and $d_k$ inherit that dependence weakly. A standard error that ignores these effects misstates the evidence: with positive autocorrelation it is too small and the test rejects too readily, with negative autocorrelation it is too large and the test loses power. Only for the optimal filter do the two standard errors agree.

The heteroskedasticity and autocorrelation consistent (HAC) estimator of Newey and West \cite{newey1987simple} provides a standard error that assumes neither independence nor a constant variance. The word consistent in HAC refers here to the estimator, which converges to the variance it estimates as $N$ grows, not to the filter. It rests on two conditions. The sequence must be strongly mixing, that is, the dependence between $d_k$ and $d_{k+j}$ must vanish as the separation $j$ grows \cite{doukhan1994mixing}, which makes the autocovariances $\gamma_\ell = \Cov(d_k, d_{k+\ell})$ summable. The sequence must also have moments of order higher than two bounded uniformly in $k$ \cite{hansen1992consistent,de2000strong}; this condition takes the place of weak stationarity, which $\{d_k\}$ does not satisfy because $\Var(d_k) = 2/m_k$ changes with the sensor, and it holds here because $d_k$ is a quadratic form of Gaussian innovations and has finite moments of every order. Under these conditions the sample mean of a dependent sequence obeys a central limit theorem \cite{herrndorf1984functional}: for large $N$, $\Var(\bar{d}) \simeq \Omega/N$, where the long-run variance $\Omega = \sum_{\ell=-\infty}^{\infty}\gamma_\ell$ adds the variance of $d_k$ and all its cross-covariances; with sensors of different dimension, each $\gamma_\ell$ is the average over the sequence of the covariances at lag $\ell$. The estimator truncates the sum at a bandwidth $L$ and weights the sample autocovariances with the Bartlett kernel,
\begin{equation}
  \hat{\Omega} = \hat{\gamma}_0 + 2\sum_{\ell=1}^{L}\Bigl(1 - \frac{\ell}{L+1}\Bigr)\hat{\gamma}_\ell, \qquad \hat{\gamma}_\ell = \frac{1}{N}\sum_{k=1}^{N-\ell}\bigl(d_k - \bar{d}\bigr)\bigl(d_{k+\ell} - \bar{d}\bigr). \label{eq:nw}
\end{equation}
The Bartlett weights decrease linearly with the lag and so downweight the high-lag autocovariances, which are estimated from fewer products; they also keep $\hat{\Omega}$ non-negative \cite{newey1987simple}. The subtraction of $\bar{d}$ in $\hat{\gamma}_\ell$ centres the sequence, whose mean is not zero when $H_0$ fails. The bandwidth follows the fixed rule $L = \lfloor 4(N/100)^{2/9}\rfloor$, which is the preliminary lag of the automatic bandwidth selection of Newey and West \cite{newey1994automatic}, used here as a rule of thumb (for $N = 100$ and $N = 1000$ it gives $L = 4$ and $L = 6$). A data-driven bandwidth, such as the automatic selection of Newey and West \cite{newey1994automatic}, is left for future work. When the $d_k$ are uncorrelated, $\hat{\gamma}_\ell \simeq 0$ for $\ell \geq 1$, and $\hat{\Omega}$ reduces to the sample variance.

\subsection{Test statistic and decision}\label{sec:decision}

The HAC-DANIS statistic is
\begin{equation}
  \Thac = \frac{\bar{d}}{\sqrt{\hat{\Omega}/N}}, \label{eq:thac}
\end{equation}
and $\Thac \xrightarrow{d} \mathcal{N}(0,1)$ under $H_0$ as $N \to \infty$, by the central limit theorem of \cite{herrndorf1984functional} and the convergence of $\hat{\Omega}$ \cite{hansen1992consistent,de2000strong}. When the filter is miscalibrated the mean of $d_k$ is $\mu_k \neq 0$ and $\bar{d}$ estimates $\mu_N$ (Section~\ref{sec:danis}). The statistic then grows as $\sqrt{N}\mu_N$, which is what drives the rejection, and the interval below covers $\mu_N$. If $\mu_k$ is the same at every epoch, the centred ratio $\sqrt{N}(\bar{d} - \mu_N)/\sqrt{\hat{\Omega}}$ is still asymptotically standard normal and the interval has its nominal coverage. If $\mu_k$ varies along the run the centring by the single value $\bar{d}$ in Eq.~\eqref{eq:nw} leaves the variation of $\mu_k$ about $\mu_N$ inside the sample autocovariances. The standard error is then conservative: the interval is wider than its nominal coverage requires and the test loses some power, while its level under $H_0$, where $\mu_k = 0$ at every epoch, is unaffected. For a two-sided test at significance level $\alpha$, $H_0$ is rejected when $|\Thac| > z_{1-\alpha/2}$, the standard normal quantile; we use $\alpha = 0.05$ throughout. The sign of $\bar{d}$ gives the direction of the miscalibration, and the interval $\bar{d} \pm z_{1-\alpha/2}\sqrt{\hat{\Omega}/N}$ gives the size of the inconsistency. Algorithm~\ref{alg:hacdanis} lists the workflow from the innovations to the decision.

\begin{algorithm}[t]
\caption{The HAC-DANIS test.}
\label{alg:hacdanis}
\begin{algorithmic}[1]
\Require innovations $\boldsymbol{\nu}_k$, innovation covariances $\boldsymbol{S}_k$ and dimensions $m_k$ for $k = 1, \dots, N$, and the significance level $\alpha$
\Ensure decision on $H_0$, with $\bar{d}$, its standard error, the $p$-value and the interval for the mean of $d_k$
\For{$k = 1, \dots, N$}
  \State $\NIS_k \gets \boldsymbol{\nu}_k^{\top}\boldsymbol{S}_k^{-1}\boldsymbol{\nu}_k$, \quad $d_k \gets \NIS_k/m_k - 1$ \Comment{Eqs.~\eqref{eq:nis} and \eqref{eq:dk}}
\EndFor
\State $\bar{d} \gets \frac{1}{N}\sum_{k=1}^{N} d_k$ \Comment{Eq.~\eqref{eq:dbar}}
\State $L \gets \lfloor 4(N/100)^{2/9}\rfloor$ \Comment{bandwidth}
\For{$\ell = 0, \dots, L$}
  \State $\hat{\gamma}_\ell \gets \frac{1}{N}\sum_{k=1}^{N-\ell}(d_k - \bar{d})(d_{k+\ell} - \bar{d})$ \Comment{Eq.~\eqref{eq:nw}}
\EndFor
\State $\hat{\Omega} \gets \hat{\gamma}_0 + 2\sum_{\ell=1}^{L}\bigl(1 - \tfrac{\ell}{L+1}\bigr)\hat{\gamma}_\ell$ \Comment{Eq.~\eqref{eq:nw}}
\State $\Thac \gets \bar{d}\big/\sqrt{\hat{\Omega}/N}$, \quad $p \gets 2\bigl[1 - \Phi(|\Thac|)\bigr]$ \Comment{Eq.~\eqref{eq:thac}, $\Phi$ standard normal}
\State reject $H_0$ if $|\Thac| > z_{1-\alpha/2}$, and read the direction of the miscalibration from the sign of $\bar{d}$
\State \Return $\bar{d}$, $\sqrt{\hat{\Omega}/N}$, $p$, $\bar{d} \pm z_{1-\alpha/2}\sqrt{\hat{\Omega}/N}$
\end{algorithmic}
\end{algorithm}

The statistic can be computed on the run, as the measurements arrive. Evaluating $\bar{d}$ and $\hat{\Omega}$ on the first $n$ measurements, with the bandwidth evaluated at $n$, gives a cumulative curve $\bar{d}(n)$. During the first tens of measurements the curve fluctuates widely, because the central limit approximation needs data and $\hat{\Omega}$ rests on few measurements; the decisions in Section~\ref{sec:results} use the complete simulation.

\subsection{Autocorrelation function of the normalised sequence}\label{sec:acf}

The HAC estimator accounts for the autocorrelation of $\{d_k\}$ without describing it. The sample autocorrelation function (ACF),
\begin{equation}
  \hat{\rho}_\ell = \hat{\gamma}_\ell/\hat{\gamma}_0, \label{eq:acf}
\end{equation}
shows that structure lag by lag. For uncorrelated $d_k$ about $95\%$ of the values fall within $\pm 1.96/\sqrt{N}$, so values outside this band at the first lags indicate dependence. Whiteness tests examine the autocorrelation of the innovation vectors of one sensor under a fixed measurement model \cite{mehra1970identification}; the ACF of $\{d_k\}$ covers all sensors in one scalar series, and its shape characterises the mistuning regime. For a well-calibrated filter ($\bar{d} \simeq 0$) the dependence that remains is weak (Section~\ref{sec:hac}), so the sequence is close to white and most lags lie inside the band. For an overconfident filter ($\bar{d} > 0$) the autocorrelation is strongly positive at short lags and decays quickly, which we read as the effect of a small gain: the filter corrects the state little at each update, so consecutive prediction errors share most of their content. For an overconservative filter ($\bar{d} < 0$) the autocorrelation is positive at short lags and turns negative at longer ones, which we attribute to the large gain: each prediction starts essentially from the previous measurement, so the dependence follows the tracking passes, strong within a pass and changing sign across passes. Section~\ref{sec:acfresults} shows these three patterns in the Starlette scenarios. The gain mechanism is a hypothesis, and the derivation of the autocorrelation of $d_k$ under miscalibration is left for future work.
 \section{Simulation setup}\label{sec:setup}

Three process noise configurations of known character are applied to four simulated satellites in contrasting dynamical regimes, and HAC-DANIS is evaluated on each of the twelve scenarios.

\subsection{Target satellites and dynamical models}\label{sec:satellites}

Table~\ref{tab:satellites} lists the four satellites. Starlette (NORAD 7646) is a passive geodetic sphere at about 810~km altitude, where atmospheric drag is small and the dynamics are dominated by gravitational perturbations. Envisat (NORAD 27386) flies at a similar altitude (about 770~km, Sun-synchronous) and is a much larger spacecraft, but its area-to-mass ratio is of the same order as Starlette's, $1.00$ against $1.14$ in units of $10^{-3}$~m$^2$/kg (Table~\ref{tab:satellites}), so its non-gravitational accelerations are comparable. What sets it apart is its orbit and its visibility from the sensors (Section~\ref{sec:estimation}). The International Space Station (ISS, NORAD 25544, about 420~km) and the Tiangong space station (NORAD 48274, about 390~km) fly where the atmospheric density is roughly two orders of magnitude larger, so drag dominates the perturbations and the error of the modelled density dominates the model error.

\begin{table}[t]
\centering
\caption{Satellite parameters used to generate the reference trajectories: altitude, mass, drag and solar radiation pressure (SRP) areas (drag / SRP), area-to-mass ratio $A/m$ with the drag area in $10^{-3}$~m$^2$/kg, drag coefficient $C_D$ and reflectivity coefficient $C_R$. The rows labelled filter give the mass, areas and initial force coefficients used by the UKF.}
\label{tab:satellites}
\small\setlength{\tabcolsep}{4pt}
\begin{tabular}{lrrrrrr}
\toprule
Satellite & Altitude (km) & Mass (kg) & Area (m$^2$) & $A/m$ & $C_D$ & $C_R$ \\

\midrule
Starlette & 810 & 46.6 & 0.053 / 0.053 & 1.14 & 2.2 & 1.8 \\
\quad\textit{filter} & & 46.57 & 0.052 / 0.052 & & 2.1 & 1.86 \\
Envisat & 770 & 8\,000 & 8.0 / 5.0 & 1.00 & 2.3 & 1.5 \\
\quad\textit{filter} & & 8\,000.5 & 8.25 / 5.1 & & 2.35 & 1.53 \\
ISS & 420 & 450\,000 & 350 / 280 & 0.78 & 2.2 & 1.4 \\
\quad\textit{filter} & & 450\,002 & 352.5 / 285.2 & & 2.05 & 1.7 \\
Tiangong & 390 & 66\,000 & 85 / 65 & 1.29 & 2.2 & 1.3 \\
\quad\textit{filter} & & 66\,005 & 84.1 / 66.8 & & 2.25 & 1.55 \\
\bottomrule
\end{tabular}
\end{table}

The reference trajectories are propagated with a high-fidelity model: a $150\times150$ geopotential, the NRLMSISE-00 atmosphere for drag, the gravity of the Sun, Moon, Venus, Mars and Jupiter, SRP, Earth albedo, relativistic corrections, solid tides and $30\times30$ ocean tides, with a position tolerance of $10^{-9}$~m. The UKF uses a reduced model: a $50\times50$ geopotential, the DTM2000 atmosphere, the gravity of the Sun and Moon, SRP, solid tides and $5\times5$ ocean tides, with a position tolerance of $10^{-5}$~m. Its spacecraft parameters (mass, cross-sectional areas and initial force coefficients) are slightly perturbed with respect to the reference values (Table~\ref{tab:satellites}, rows labelled filter). The estimator therefore works with an imperfect dynamical model and an imperfect description of the spacecraft, as in operational orbit determination. All trajectories are computed with the Orekit astrodynamics library, version 13.1, with the Hipparchus 4.0.3 mathematics library beneath it.

\subsection{Sensor network}\label{sec:sensors}

Observations are simulated for three sensors similar to those of the S3T network \cite{gomez2017description}, the Spanish space surveillance and tracking network introduced in Section~\ref{sec:intro}: a laser ranging station, an optical telescope and a surveillance radar (Table~\ref{tab:sensors}). Their measurement dimensions, $m_k \in \{1, 2, 4\}$, form the heterogeneous case that motivates the normalisation of Section~\ref{sec:danis}. The measurements are corrupted with zero-mean Gaussian noise of the standard deviations in Table~\ref{tab:sensors}, and the filter uses the same values in $\boldsymbol{R}_k$. The measurement noise model is therefore exact in every run, and any departure from consistency comes from the process noise. Within a pass the laser and the telescope deliver one measurement every $25$~s and the radar one every $12$~s, while the passes of a sensor are separated by gaps of hours (Section~\ref{sec:estimation}).

\begin{table}[t]
\centering
\caption{Characteristics of the simulated sensors.}
\label{tab:sensors}
\begin{tabular}{llcl}
\toprule
Sensor & Observables & $m_k$ & Noise ($1\sigma$) \\
\midrule
Laser & Range & 1 & 0.01~m \\
Telescope & Right ascension, declination & 2 & 1.4~arcsec \\
Radar & Range & 4 & 7~m \\
 & Range rate & & 0.4~m/s \\
 & Azimuth, elevation & & 0.35$^\circ$ \\
\bottomrule
\end{tabular}
\end{table}

\subsection{Process noise configurations}\label{sec:qconfigs}

Three process noise configurations are run for each satellite, all with the polynomials of Eq.~\eqref{eq:Qpoly} and the coefficients of Table~\ref{tab:qcoef}, in which a single constant $C$ sets the level of the noise. In the overconfident configuration $\boldsymbol{Q}$ lies below the level needed to absorb the dynamical model errors, so the filter underestimates its uncertainty, and HAC-DANIS should give $\bar{d} > 0$ with a large positive $\Thac$. In the overconservative configuration $\boldsymbol{Q}$ lies above that level, and HAC-DANIS should give $\bar{d} < 0$ with a negative $\Thac$. In the well-calibrated configuration $C$ was varied between the two miscalibrated values and the value that brings $\bar{d}$ closest to zero over the observation period was retained to obtain the well-calibrated cases. These four scenarios therefore show the filter at a process noise the test accepts. The constant $C$ is an exponent on the time unit $\tau = 86\,400$~s, so it scales the position and velocity coefficients differently: relative to $C = 1$, $a_{2,x}$ is divided by $\tau^{2(C-1)}$ and $b_{1,x}$ by $\tau^{C-1}$. The coefficients and $C$ values are shown in Table~\ref{tab:qcoef}.

\begin{table}[t]
\centering
\caption{Coefficients of the process noise polynomials of Eq.~\eqref{eq:Qpoly}: $a_{1,x} = 0$, $a_{2,x} = A_x/\tau^{2C}$ and $b_{1,x} = B_x/\tau^{C}$ with $\tau = 86\,400$~s and $\Delta t$ in seconds, so that for $C = 1$ the along-track noise accumulated over one day is $800$~m in position and $5$~m/s in velocity. $C = 1.2$ in the overconservative configuration, $C = 1.4$ in the overconfident one and between $1.25$ and $1.35$, depending on the satellite, in the well-calibrated one.}
\label{tab:qcoef}
\begin{tabular}{lcccc}
\toprule
Axis & $a_{0,x}$ (m) & $A_x$ (m) & $b_{0,x}$ (m/s) & $B_x$ (m/s) \\
\midrule
$T$ & $10^{-3}$ & 800 & $10^{-6}$ & 5 \\
$N$ & $2\cdot10^{-4}$ & 150 & $2\cdot10^{-7}$ & 1 \\
$W$ & $5\cdot10^{-5}$ & 75 & $5\cdot10^{-8}$ & 0.5 \\
\bottomrule
\end{tabular}
\end{table}

\subsection{Estimation configuration}\label{sec:estimation}

Each scenario consists of a two-day observation period, over which the measurements are processed, followed by three days of free propagation, over which the error of the final estimate is followed. Table~\ref{tab:measurements} gives the number of measurements available for each satellite. The three filters of a satellite, one per process noise configuration, use the same measurements, so the differences between them come from $\boldsymbol{Q}$ alone. The telescope had no visibility of Envisat over its observation period, so that scenario contains laser and radar measurements only. With these $N$, the bandwidth rule of Section~\ref{sec:hac} gives $L = 4$ for Envisat, the ISS and Tiangong and $L = 5$ for Starlette. The filter is the unscented Kalman estimator of Orekit, built on the Hipparchus filter of Section~\ref{sec:ukfsteps}, with the sigma-point parameters at their default values $\alpha = 0.5$, $\beta = 2$ and $\kappa = 0$. The initial state and covariance come from the initial orbit determination method of Montilla et al.\ \cite{montilla2024single}, whose position standard deviations are of the order of a kilometre. The drag coefficient $C_D$ and the reflectivity coefficient $C_R$ are estimated together with the orbital state, starting from approximate values that differ from the reference ones (Section~\ref{sec:satellites}) and with a constant process noise of standard deviation about $10^{-4}$ per step, small enough only to let the estimates move. No trajectory contains manoeuvres.

\begin{table}[t]
\centering
\caption{Number of measurements available for each satellite, by sensor, and total $N$.}
\label{tab:measurements}
\begin{tabular}{lrrrr}
\toprule
Satellite & Laser & Telescope & Radar & $N$ \\
\midrule
Starlette & 336 & 126 & 59 & 521 \\
Envisat   & 162 & 0   & 50 & 212 \\
ISS       & 74  & 35  & 18 & 127 \\
Tiangong  & 101 & 55  & 23 & 179 \\
\bottomrule
\end{tabular}
\end{table}
 \section{Results}\label{sec:results}

In this section the twelve scenarios are analysed, first through the estimation performance of the filters, then through the HAC-DANIS decisions, the autocorrelation structure of the normalised sequence and, finally, the variation of the HAC standard error across configurations.

\subsection{Estimation performance}\label{sec:performance}

Table~\ref{tab:performance} summarises the estimation performance of the twelve scenarios. Its two groups of columns are evaluated at different instants. The position error, summarised by its median, its 95th percentile and its root mean square (RMS), is evaluated at every time step of the run, including the gaps between sensor passes and the three days of propagation after the last measurement. The reported uncertainty comes from the posterior covariance after each measurement update, the only covariance recorded. The two groups therefore cannot be directly compared with each other; each is compared across configurations.

\begin{table}[t]
\centering
\caption{Estimation performance for each satellite and process noise configuration. Position error over every time step of the run, including the three days of propagation after the last measurement. Reported uncertainty, the square root of the trace of the position covariance, after each measurement update.}
\label{tab:performance}
\small
\begin{tabular}{lrrrrr}
\toprule
 & \multicolumn{3}{c}{Position error (m)} & \multicolumn{2}{c}{Reported uncertainty (m)} \\
\cmidrule(lr){2-4} \cmidrule(lr){5-6}
Configuration & Median & 95th pct. & RMS & Median & 95th pct. \\
\midrule
\multicolumn{6}{l}{\textit{Starlette}} \\
Overconservative & 231 & 2\,559 & 966 & 14.2 & 174 \\
Overconfident & 6.3 & 37.1 & 16.8 & 0.008 & 0.015 \\
Well-calibrated & 13.1 & 26.2 & 19.0 & 4.2 & 11.4 \\
\midrule
\multicolumn{6}{l}{\textit{Envisat}} \\
Overconservative & 1\,799 & 13\,628 & 5\,236 & 95.1 & 351 \\
Overconfident & 204 & 550 & 298 & 0.066 & 1.38 \\
Well-calibrated & 7.4 & 40.2 & 18.8 & 5.5 & 14.6 \\
\midrule
\multicolumn{6}{l}{\textit{ISS}} \\
Overconservative & 208\,782 & 814\,518 & 411\,769 & 8.4 & 107 \\
Overconfident & 23.8 & 106 & 53.2 & 0.031 & 0.049 \\
Well-calibrated & 10.7 & 25.8 & 13.4 & 3.8 & 9.4 \\
\midrule
\multicolumn{6}{l}{\textit{Tiangong}} \\
Overconservative & 7\,070 & 33\,302 & 13\,125 & 6.1 & 151 \\
Overconfident & 273 & 3\,050 & 1\,331 & 0.012 & 0.064 \\
Well-calibrated & 306 & 651 & 412 & 2.2 & 89.5 \\
\bottomrule
\end{tabular}
\end{table}

The overconservative configurations give the largest position errors on every satellite, one to four orders of magnitude above the well-calibrated ones, and they report larger covariances than the well-calibrated configurations (median uncertainties of $14.2$~m against $4.2$~m for Starlette and $95.1$~m against $5.5$~m for Envisat). The inflated $\boldsymbol{Q}$ makes the filter distrust its dynamical prediction: the predicted covariance grows quickly between passes, and the large gain pulls the estimate towards each new measurement without keeping a stable solution from one pass to the next. The ISS overconservative filter reaches a median error of $209$~km. We attribute this to the exposure of the ISS to drag at $420$~km and the low number of available measurements (Table~\ref{tab:measurements}). A filter that trusts each new measurement much more than its prediction can move the estimated orbit away from the true one at an update, and the differences between the models of the filter and those of the reference trajectory, in particular the DTM2000 and NRLMSISE-00 atmospheres, then make the error grow quickly between passes and during the propagation.

The overconfident configurations show the opposite failure. Their position errors range from comparable to the well-calibrated case, for Starlette, to much larger, for Envisat, while the covariance they report is orders of magnitude smaller (median uncertainties of $0.008$~m for Starlette and $0.031$~m for ISS, against $4.2$~m and $3.8$~m in the well-calibrated case). The filter follows the spacecraft but has no reliable estimate of its own accuracy. In operations this failure is particularly dangerous, because conjunction assessment and re-entry prediction depend critically on realistic covariance estimates, and assuming a low error level when it is in fact incorrect can lead to catastrophic events.

The well-calibrated configurations report uncertainties between those of the other two configurations and give the lowest RMS error for Envisat, ISS and Tiangong. For Starlette the overconfident filter has a slightly lower RMS error ($16.8$~m against $19.0$~m) while reporting a covariance orders of magnitude smaller; the calibration of $\boldsymbol{Q}$ concerns the agreement between the reported covariance and the error more than the size of the error. Every Tiangong configuration gives errors of hundreds of metres or more over the full run, which we attribute to the propagation through the mismodelled atmosphere and the small number of measurements available during the observation period (Table~\ref{tab:measurements}). There the well-calibrated filter does not improve on the overconfident one in median error ($306$~m against $273$~m), but its reported uncertainty is much larger ($2.2$~m against $0.012$~m). These values are not a consistency check of the well-calibrated filter, since the error column is dominated by the three days of propagation while the uncertainty is that of the update epochs and they are not directly comparable.

\subsection{HAC-DANIS results}\label{sec:hacresults}

Table~\ref{tab:hacdanis} gives the HAC-DANIS statistics of the twelve scenarios. Every overconservative configuration is rejected with $\bar{d} < 0$, every overconfident configuration is rejected with $\bar{d} > 0$, and no well-calibrated configuration is rejected. The sign of $\bar{d}$ identifies the direction of the miscalibration in the eight miscalibrated cases, and the $p$-value measures the strength of the evidence. Each scenario is a single run, so these results show the test reaching the expected decision in each case.

\begin{table}[!tbp]
\centering
\caption{HAC-DANIS statistics for each satellite and process noise configuration: mean $\bar{d}$, HAC standard error (SE), two-sided $p$-value and decision at $\alpha = 0.05$. Standard errors are rounded to two significant figures and the $p$-values come from the unrounded statistics.}
\label{tab:hacdanis}
\small
\begin{tabular}{lrrrl}
\toprule
Configuration & $\bar{d}$ & HAC SE & $p$-value & Decision \\
\midrule
\multicolumn{5}{l}{\textit{Starlette}} \\
Overconservative & $-0.49$ & 0.051 & $<10^{-4}$ & Reject ($\boldsymbol{Q}$ large) \\
Overconfident & $+0.63$ & 0.18 & $3\cdot10^{-4}$ & Reject ($\boldsymbol{Q}$ small) \\
Well-calibrated & $-0.002$ & 0.061 & 0.97 & Fail to reject \\
\midrule
\multicolumn{5}{l}{\textit{Envisat}} \\
Overconservative & $-0.57$ & 0.094 & $<10^{-4}$ & Reject ($\boldsymbol{Q}$ large) \\
Overconfident & $+2.90$ & 0.78 & $2\cdot10^{-4}$ & Reject ($\boldsymbol{Q}$ small) \\
Well-calibrated & $+0.065$ & 0.10 & 0.52 & Fail to reject \\
\midrule
\multicolumn{5}{l}{\textit{ISS}} \\
Overconservative & $-0.39$ & 0.079 & $<10^{-4}$ & Reject ($\boldsymbol{Q}$ large) \\
Overconfident & $+0.197$ & 0.094 & 0.036 & Reject ($\boldsymbol{Q}$ small) \\
Well-calibrated & $+0.142$ & 0.133 & 0.29 & Fail to reject \\
\midrule
\multicolumn{5}{l}{\textit{Tiangong}} \\
Overconservative & $-0.406$ & 0.062 & $<10^{-4}$ & Reject ($\boldsymbol{Q}$ large) \\
Overconfident & $+5889$ & 2059 & 0.004 & Reject ($\boldsymbol{Q}$ small) \\
Well-calibrated & $-0.048$ & 0.051 & 0.35 & Fail to reject \\
\bottomrule
\end{tabular}
\end{table}

\begin{figure}[!tbp]
\centering
\includegraphics[width=0.70\textwidth]{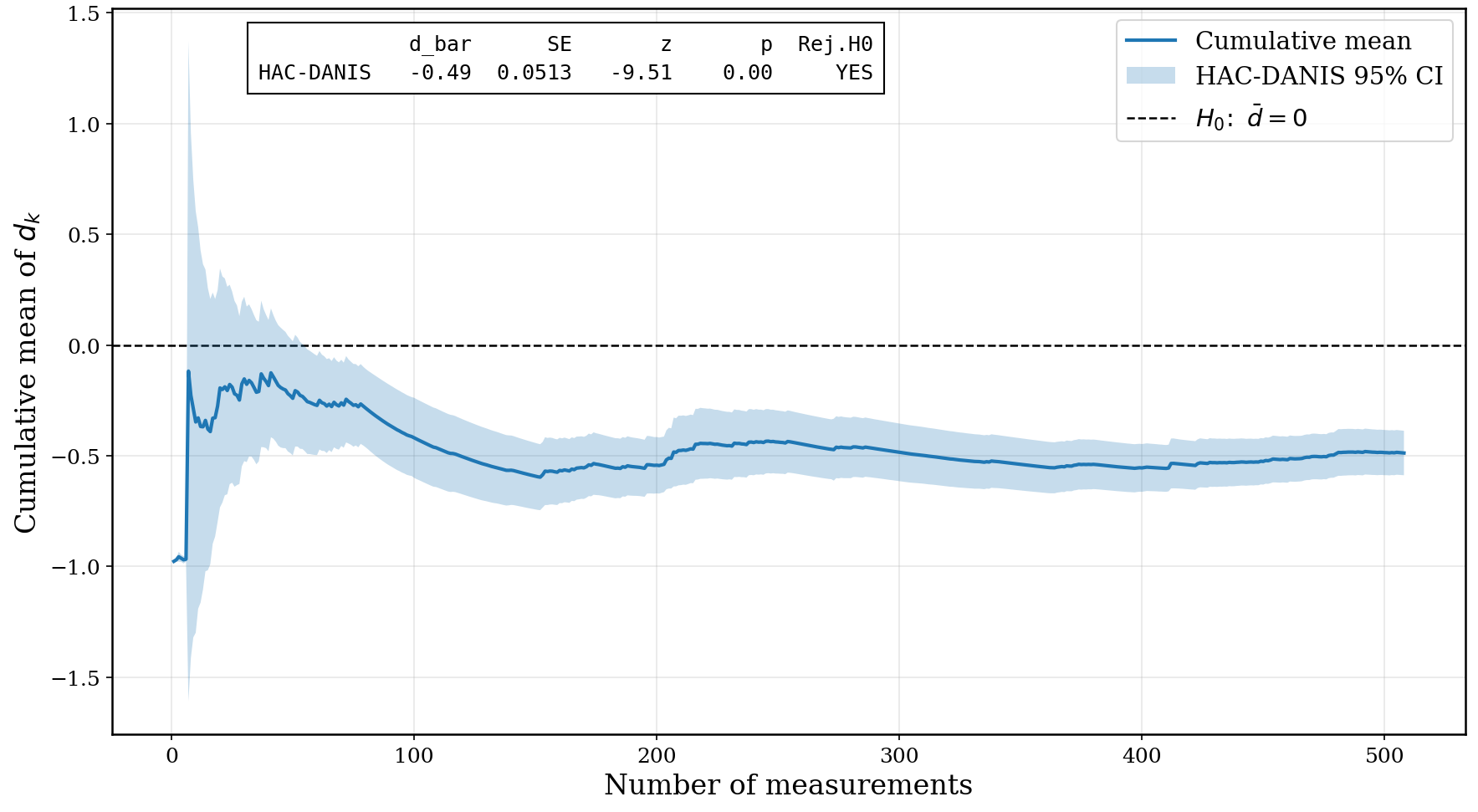}\\[1ex]
\includegraphics[width=0.70\textwidth]{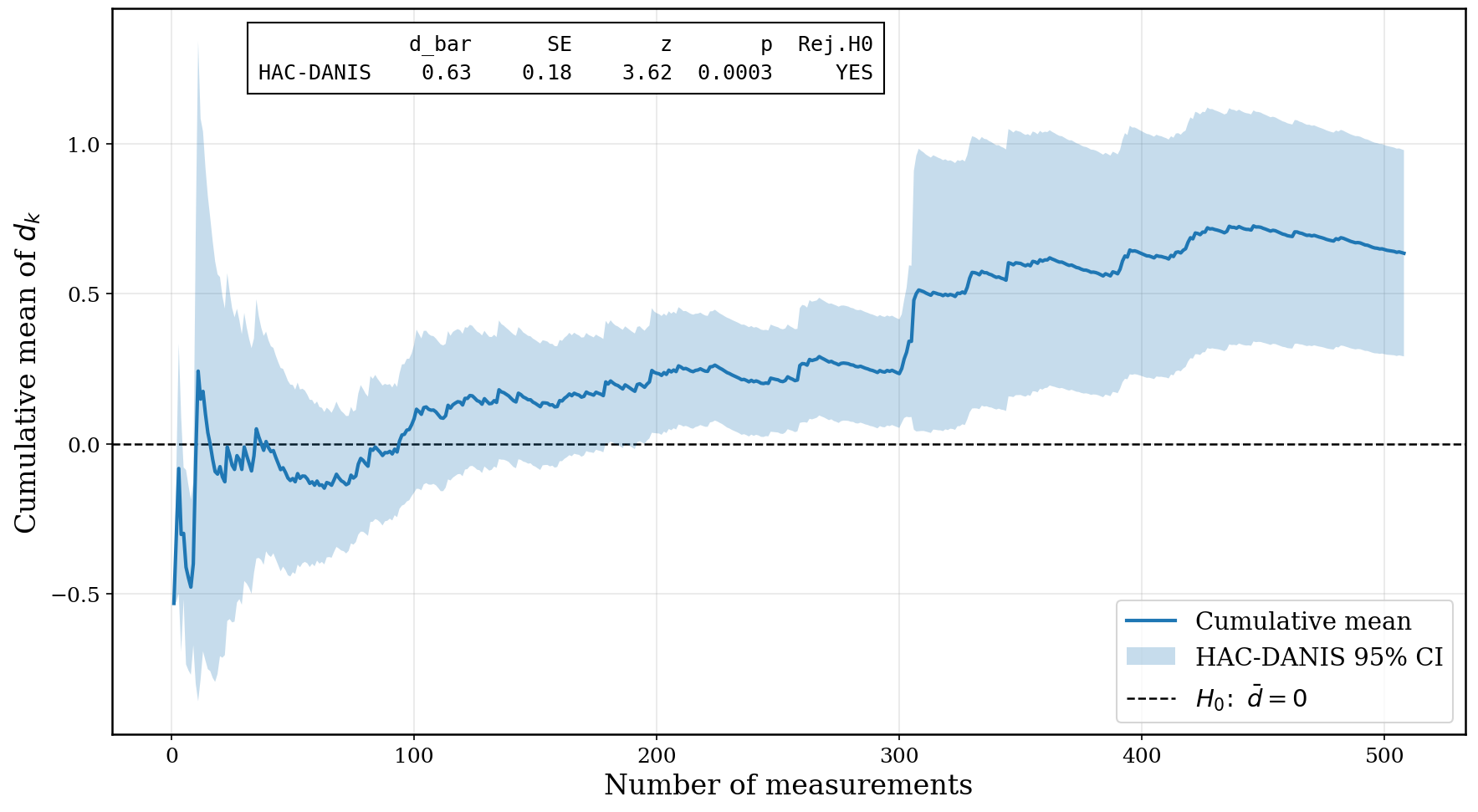}\\[1ex]
\includegraphics[width=0.70\textwidth]{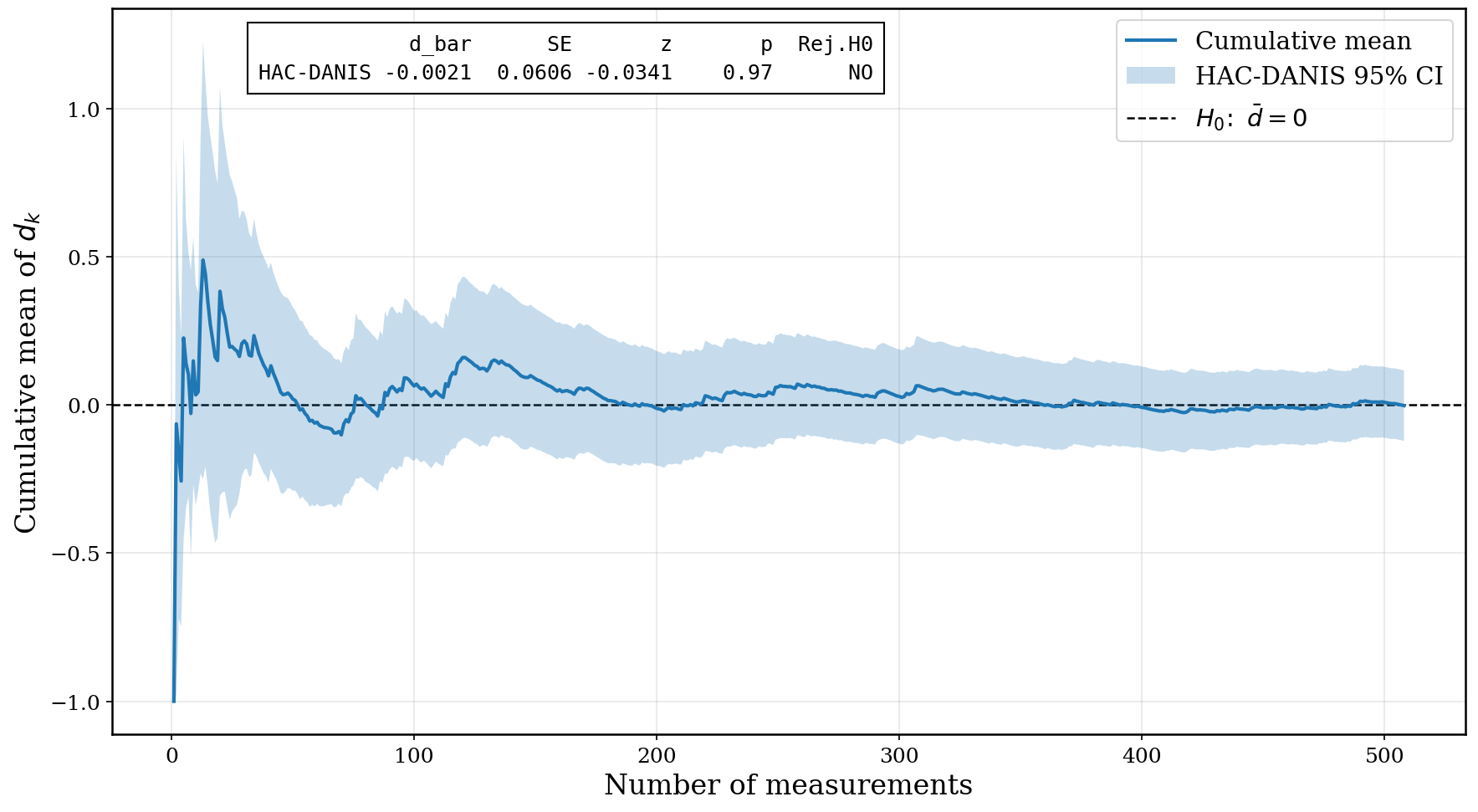}
\caption{Cumulative HAC-DANIS evolution for Starlette in the overconservative (top), overconfident (middle) and well-calibrated (bottom) configurations: cumulative mean $\bar{d}(n)$ of $d_k$ (line) and its $95\%$ HAC interval (band) against the number of processed measurements, with the reference $\bar{d} = 0$ (dashed line). The inset gives the final $\bar{d}$, the HAC standard error (SE), the statistic $\Thac$ (labelled $z$), the $p$-value and the decision on $H_0$.}
\label{fig:onrun}
\end{figure}

Fig.~\ref{fig:onrun} shows the cumulative evolution of $\bar{d}$ for Starlette. In the overconservative configuration $\bar{d}(n)$ settles at $-0.49$, and its interval excludes zero after about $50$ measurements. In the overconfident configuration $\bar{d}(n)$ grows progressively and ends at $+0.63$. In the well-calibrated configuration it fluctuates around zero and its confidence interval contains it along the whole run.

The weakest rejection is that of the ISS overconfident configuration ($p = 0.036$). Its mean, $\bar{d} = +0.197$, is close to that of the ISS well-calibrated run, $\bar{d} = +0.142$ ($p = 0.29$), and the two decisions differ because of the standard errors, $0.094$ against $0.133$. The ISS has the smallest sample of the four ($N = 127$), and these two cases show how close a rejection and a non-rejection can lie at that sample size.

The Tiangong overconfident configuration is qualitatively different from the others. Its mean $\bar{d} \simeq 5889$ indicates NIS values thousands of times their expected magnitude: the reported covariance has collapsed to a median uncertainty of $0.012$~m (Table~\ref{tab:performance}) and the innovations are enormous relative to $\boldsymbol{S}_k$. The enormous HAC standard error ($2059$) reflects the extreme variability of the $d_k$ sequence in this regime. The test still rejects $H_0$ ($p = 0.004$), so the decision is the expected one even under catastrophic miscalibration. In such cases the magnitude of $\bar{d}$ alone is sufficient to diagnose the problem.

\subsection{Autocorrelation structure across configurations}\label{sec:acfresults}

\begin{figure}[!tbp]
\centering
\includegraphics[width=\textwidth]{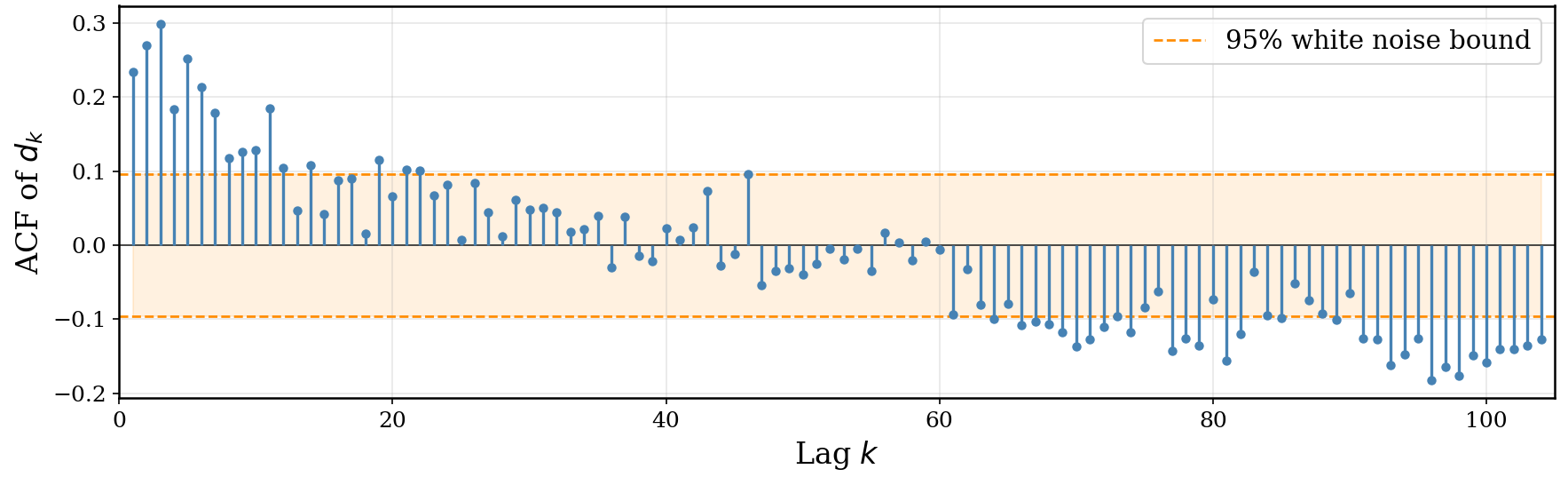}\\[1ex]
\includegraphics[width=\textwidth]{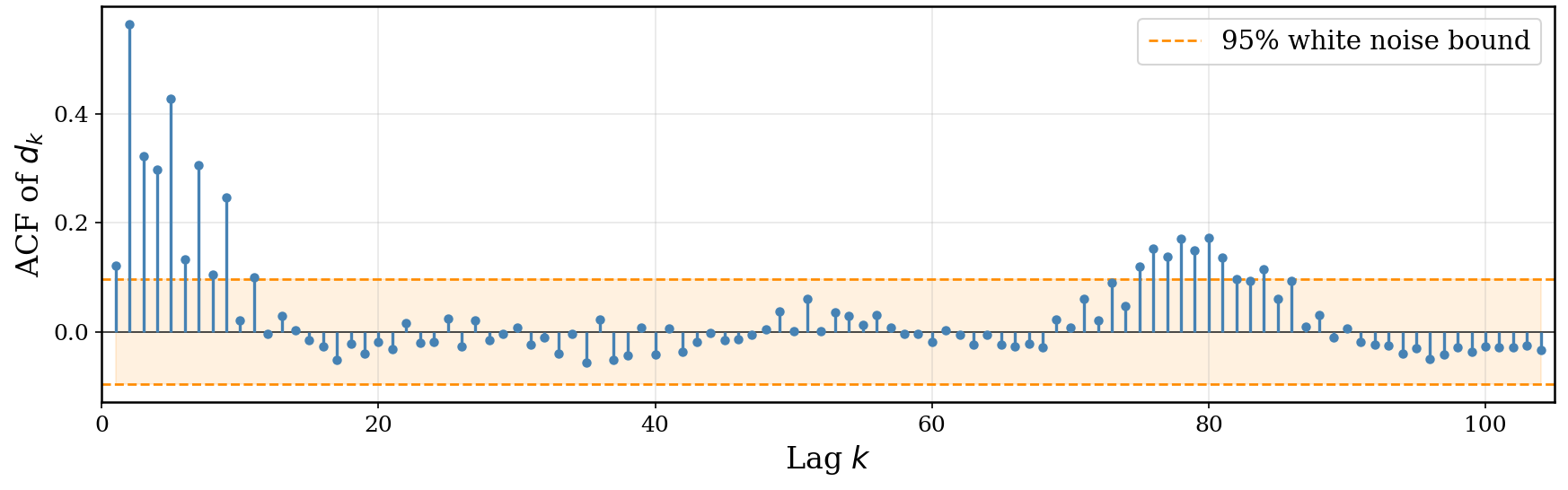}\\[1ex]
\includegraphics[width=\textwidth]{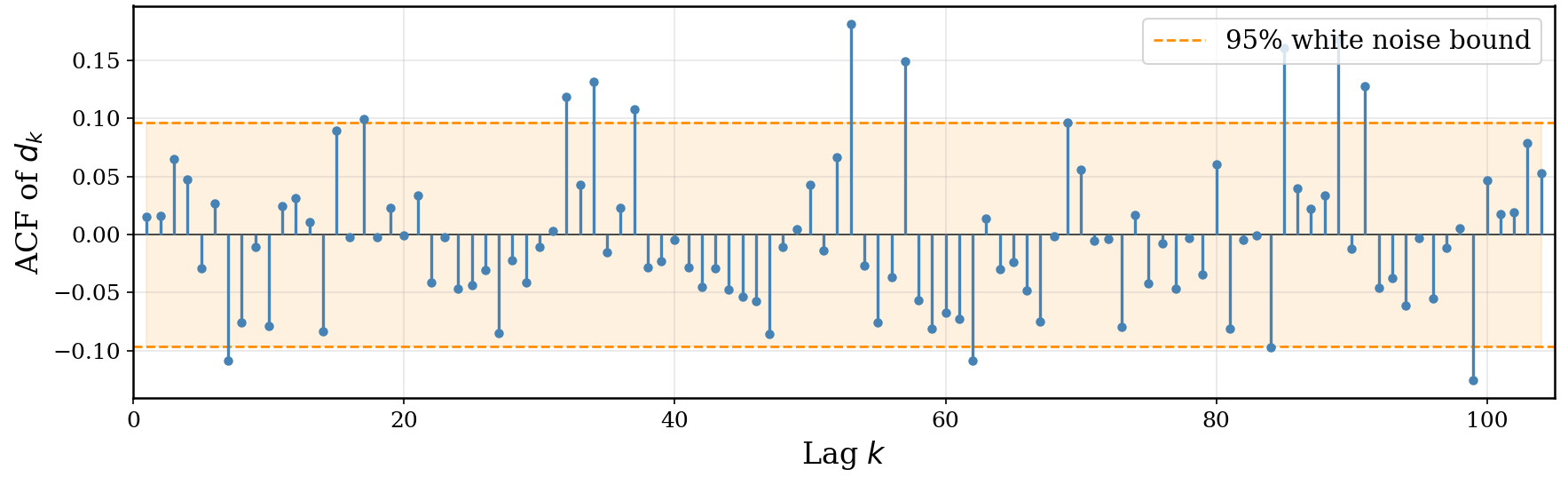}
\caption{Sample autocorrelation function of $d_k$ for Starlette in the overconservative (top), overconfident (middle) and well-calibrated (bottom) configurations, against the lag (labelled $k$) in number of measurements. The dashed lines and the shaded band mark the $95\%$ white-noise bound $\pm 1.96/\sqrt{N}$.}
\label{fig:acf}
\end{figure}

Fig.~\ref{fig:acf} shows the ACF of $\{d_k\}$ for Starlette. In the well-calibrated configuration the sequence is close to white: the autocorrelation is small at every lag and shows no structure. In the overconfident configuration the autocorrelation is strongly positive at short lags and decays quickly. We interpret this as the effect of a small gain: the filter corrects the state little at each update, so consecutive predictions share most of their error. In the overconservative configuration the autocorrelation is positive at the first lags and becomes negative at longer ones. We attribute this to the large gain: each prediction starts essentially from the previous measurement, so the dependence follows the tracking passes, strong within a pass and changing sign across passes. The three patterns are those described in Section~\ref{sec:acf}, and the derivation of their relation to the gain is left for future work.

\subsection{Variation of the HAC standard error}\label{sec:hacrole}

The variation of the HAC standard error across configurations in Table~\ref{tab:hacdanis} reflects their different autocorrelation structures. For Starlette, the standard error of the overconfident configuration ($0.18$) is about three times that of the overconservative ($0.051$) and well-calibrated ($0.061$) ones, and for Envisat the overconfident standard error ($0.78$) is nearly an order of magnitude above the other two ($0.094$ and $0.10$). Two effects contribute: the larger variance of $d_k$ when the innovations exceed their predicted size, and the positive autocorrelation at short lags of Fig.~\ref{fig:acf}, which the estimator adds through its lagged terms. A detailed analysis of the mechanisms behind these differences is left for future work.

The cumulative evolution plots (Fig.~\ref{fig:onrun}) also show the convergence dynamics. In the three Starlette configurations the first $20$ to $50$ measurements show a transient driven by the small number of measurements processed, as the central limit approximation and the HAC estimate need data. Once enough measurements have been processed, $\bar{d}(n)$ indicates the calibration regime, trending downwards, upwards or towards zero depending on the process noise configuration.
 \section{Conclusions}\label{sec:conclusions}

This paper presented the HAC-DANIS test for process noise calibration in multi-sensor orbit determination. The dimensionless statistic $d_k = \NIS_k/m_k - 1$ puts the consistency evidence of sensors of different measurement dimension on a common scale, and the Newey--West HAC estimator gives an asymptotically valid long-run variance that accounts for the heteroskedasticity and the serial autocorrelation of the normalised sequence. The result is a single statistic with a fixed reference of zero whose sign diagnoses the direction of the miscalibration and whose mean $\bar{d}$ measures the size of the inconsistency. It needs no ground truth, it tolerates the serial dependence of the sequence without a model of it, and it can be followed as the measurements arrive. It examines the hypothesis that the classical NIS test examines, the mean level of the normalised innovations, with equal weight for every measurement and without the assumption of white innovations on which the classical test rests. That hypothesis is a necessary condition for consistency, not a sufficient one.

The test was applied to a UKF processing simulated laser, telescope and radar observations of four LEO satellites, Starlette, Envisat, the ISS and Tiangong, from about $810$~km to about $390$~km. The eight miscalibrated configurations were rejected with the sign of their miscalibration. The decision was also the expected one at the two extremes of miscalibration, the covariance collapse of the Tiangong overconfident filter ($\bar{d} \simeq 5889$), where the size of $\bar{d}$ alone identifies the failure, and the near-threshold ISS overconfident case ($p = 0.036$). The estimation results showed the importance of the diagnosis: the overconfident filters tracked the satellites with errors comparable to or much larger than those of the well-calibrated filters while reporting covariances orders of magnitude smaller, which heavily affects conjunction assessment and re-entry prediction. The autocorrelation function of the normalised sequence was close to white for the well-calibrated filter, strongly positive at short lags for the overconfident one, and positive and then negative, following the tracking passes, for the overconservative one, a pattern we attribute to the gain of the filter (Section~\ref{sec:acf}).

These results come from one simulated run per scenario with an exact measurement noise model. Beyond filter tuning, a well-calibrated process noise is also important for subsequent analyses: departures from $H_0$ along the run could signal non-cooperative manoeuvres or abrupt changes in atmospheric density, a persistent bias in $\bar{d}$ could indicate systematic drag mismodelling, and consistent covariances are the basis of conjunction assessment and of propagation through measurement gaps.

Several directions for future work follow from these results. First, an empirical study of the size and the power of the test with repeated runs. Second, the derivation of the autocorrelation of $\{d_k\}$ under miscalibration and of its relation to the gain, together with the structure of the mean of $d_k$ along the run, which would explain the three patterns observed. Third, a data-driven bandwidth in place of the fixed rule. Fourth, as the number of measurements in the cumulative statistic grows, each new observation changes $\bar{d}$ less and the test loses sensitivity to abrupt changes in consistency. A rolling-window variant of HAC-DANIS would keep that sensitivity and offers a basis for manoeuvre detection in non-cooperative tracking. Fifth, a measurement bias or an error in the measurement noise shifts $\bar{d}$ as a process noise error does, and separating the two needs per-sensor diagnostics. Finally, coupling HAC-DANIS with online adaptive tuning of $\boldsymbol{Q}$ would close the loop between diagnosis and correction and allow autonomous filter calibration in multi-sensor networks.

\section*{Declaration of competing interest}
The authors declare that they have no known competing financial interests or personal relationships that could have appeared to influence the work reported in this paper.

\section*{Funding}
This work was supported by grant PID2023-147623OB-I00 funded by MICIU/AEI/10.13039/501100011033 and by ``ERDF A way of making Europe''; by the Programa Operativo PPIT-FEDER Andaluc\'ia 2021--2027 [grant number SOL2024-31596]; and by the Consejer\'ia de Universidad, Investigaci\'on e Innovaci\'on of the Junta de Andaluc\'ia, Estrategia de I+D+I de Andaluc\'ia (EIDIA), Horizonte 2027 [grant number USE-26084-W].

\section*{Declaration of generative AI and AI-assisted technologies in the manuscript preparation process}
The authors would like to acknowledge the use of Claude by Anthropic (models Claude Fable 5.1 and Claude Opus 5.5) for language improvement in the preparation of this manuscript. After using this tool, the authors reviewed and edited the content as needed and take full responsibility for the content of the publication.
 
\bibliographystyle{elsarticle-num}

\end{document}